\documentclass[10pt]{article}
\usepackage{graphicx}

\def\Title#1{\begin{center} {\Large #1 } \end{center}}
\def\Author#1{\begin{center}{ \sc #1} \end{center}}
\def\Address#1{\begin{center}{ \it #1} \end{center}}

\newcommand\pubblock{\rightline{\begin{tabular}{l} Proceedings of the Second Annual LHCP\\ \pubnumber\\
         \pubdate  \end{tabular}}}

\newenvironment{Abstract}{\begin{quotation} \begin{center} 
             \large ABSTRACT \end{center}\bigskip 
      \begin{center}\begin{large}}{\end{large}\end{center} \end{quotation}}

\newenvironment{Presented}{\begin{quotation} \begin{center} 
             PRESENTED AT\end{center}\bigskip 
      \begin{center}\begin{large}}{\end{large}\end{center} \end{quotation}}

\def\Acknowledgements{\bigskip  \bigskip \begin{center} \begin{large}
             \bf ACKNOWLEDGEMENTS \end{large}\end{center}}

\def\beq{\begin{equation}}
\def\eeq#1{\label{#1}\end{equation}}
\def\eeqn{\end{equation}}

\def\beqa{\begin{eqnarray}}
\def\eeqa#1{\label{#1}\end{eqnarray}}
\def\eeqan{\end{eqnarray}}

\def\sst{\scriptscriptstyle}

\let\bar=\overbar

\def\Dslash{\not{\hbox{\kern-4pt $D$}}}
\def\dslash{\not{\hbox{\kern-2pt $\del$}}}

\def\msb{{\bar{\ssstyle M \kern -1pt S}}}

\usepackage{color}
\usepackage{amssymb}
\usepackage{array}
\usepackage{xspace}
\usepackage{enumerate}
\usepackage{subfigure}
\usepackage{lineno}
\usepackage{atlasphysics}
\usepackage{cite}

\newcommand\pubnumber{ ATL-PHYS-PROC-2014-105 }

\newcommand\pubdate{\today}

\def\affiliation{
On behalf of the ATLAS collaboration, \\
International Center for Elementary Particle Physics\\
The University of Tokyo, Tokyo 113-0033, Japan }

\def\support{\footnote{Work supported by  JSPS Grant-in-Aid for Young Scientists (A) 26707010}}

\begin{document}

\large
\begin{titlepage}
\pubblock


\Title{ Search for long-lived particles with the ATLAS detector}
\vfill

\Author{ Shimpei Yamamoto \support}
\Address{\affiliation}
\vfill
\begin{Abstract}

Several scenarios beyond the Standard Model predict heavy long-lived particles as a result of a kinematic constraint, a conserved quantum number or a weak coupling.
Such particles are possibly identified based on the detection through abnormal energy losses, appearing or disappearing tracks, displaced vertices, lepton-jet signatures, long time-of-flight  or late calorimetric energy deposits.
This contribution summarizes recent results of searches for heavy long-lived particles with the ATLAS detector using $pp$ collision data at a center of mass energy of 8~\TeV.

\end{Abstract}
\vfill

\begin{Presented}
The Second Annual Conference\\
 on Large Hadron Collider Physics \\
Columbia University, New York, U.S.A \\ 
June 2-7, 2014
\end{Presented}
\vfill
\end{titlepage}
\def\thefootnote{\fnsymbol{footnote}}
\setcounter{footnote}{0}
%

\normalsize 


\section{Introduction}

Heavy long-lived particles are predicted by various extensions of the Standard Model (SM), such as supersymmetry (SUSY),  theories with extra dimensions, scenarios with a hidden valley, and various others. 
After the absence of any signal beyond SM in generic searches at the Large Hadron Collider (LHC) up to now, exploring long-lived particles is being particularly of importance since they could  give a possibility to evade the current constraints of SM extensions.
These particles provide unique experimental signatures, and are heavily exploited in the ATLAS physics program.
Given unusual signatures that are often discarded in collision events, the analyses employ special techniques to maintain the information and reveal their characteristics.

In particular, null-observation of squarks and gluinos, in classic searches with events involving multiple energetic jets plus large missing transverse energy (\MET), is impacting SUSY extensions that has attracted a lot of work during the past decades.
There are, however, a number of theoretical frameworks that suggest different signal signatures (missed or hardly detectable) at the LHC; they could also retain the benefits of the unification of gauge groups at a high scale and the successful dark matter prediction (even without $R$-pairty conservation).
This article summarizes the following SUSY long-lived particle searches that utilize the $pp$ collision data at $\rts = 8 \TeV$ collected in 2012:
\begin{itemize}
\item Gluinos being stable and decaying late in out of bunch crossings~\cite{Aad:2013gva},
\item Charginos with a significant lifetime~\cite{Aad:2013yna},
\item Stable sleptons traversing the detector~\cite{TheATLAScollaboration:2013qha},
\item Displaced decays of heavy particles~\cite{TheATLAScollaboration:2013yia}.
\end{itemize}

\section{Signatures and searches}
\subsection{Energetic jets in out-of-time events}

The Higgs boson measured at $125~\GeV$~\cite{Aad:2012tfa,Chatrchyan:2012ufa} motivates heavy scalar scenarios where the squarks and sleptons are rendered heavy  while gauginos may still be light (at the TeV scale or below) enough to be produced at the LHC energies.
Split SUSY~\cite{ArkaniHamed:2004fb,ArkaniHamed:2004yi} is a possibility that compromises with the current constraints.
In such a scenario, gluinos decay via internal heavy squark lines, leading to their significantly long lifetimes and hadronic bound states (so-called $R$-hadrons).
Massive $R$-hadrons could be detected as low-$\beta$ particles leaving activities due to large energy loss.
Especially when they are produced nearly at the production threshold in $pp$ collisions at the LHC energies, they could come to rest inside the calorimeter and decay after a significant time, which leads to a unique signature of energetic jets delayed with respect to the collision timing.
The search for such a signature is performed by using the full $7$ and $8~\TeV$ $pp$ collision data, and adopting a dedicated trigger that selects events containing energetic jets in empty bunch crossings of the LHC.
This approach gives sensitivity to $R$-hadrons approximately with the lifetime range from  $10^{-6}$ to $10^8$ seconds.

Background events mimicking the signal signature predominantly originate from cosmic rays and upstream beam-halo interactions.
The early part  of the data with low luminosity is used to evaluate the cosmic-ray background, while a control data sample of beam-halo events is collected in unpaired bunch crossings in which only one of the two beams contains a proton bunch.
Given the leading jet energy distribution for the candidate events, no excess above the background expectation is found.

\begin{figure}[htb]
  \centering
  \includegraphics[width=0.50\textwidth]{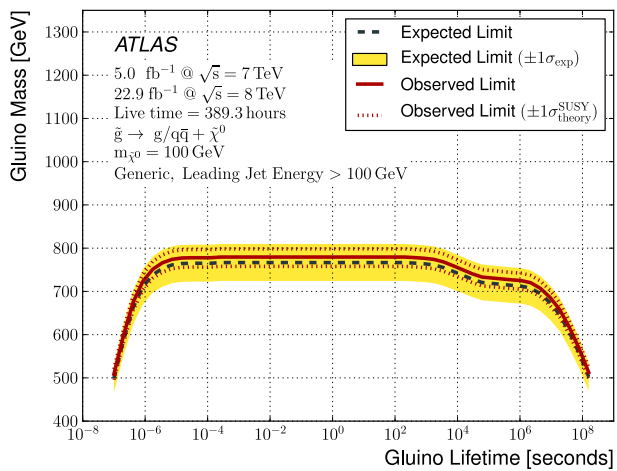}
  \caption{Limits on gluino mass versus its lifetime with $R$-hadron lifetimes in the plateau acceptance region between $10^{-5}$ and $10^3$ seconds.}
  \label{fig:figure1}
\end{figure}

Using the yield of events with the jet energy $>300~\GeV$, the upper limit on the cross section of the pair-gluino production is set as a function of the gluino mass for a given range of lifetime values,  as shown in Fig.~\ref{fig:figure1}.
The result is also interpreted for scenarios with squark $R$-hadrons or a small difference between the gluino and neutralino masses. 

\subsection{Dissapearing tracks}
A small mass difference  between the lightest chargino and neutralino ($\Delta m_{\tilde{\chi}_1}$) can occur in scenarios where they are dominantly wino or higgsino, as in anomaly-mediated SUSY breaking (AMSB)~\cite{Giudice:1998xp,Randall:1998uk} or so-called natural SUSY models: these retain a good candidate for dark matter, and could in principle accommodate the measured Higgs mass.
One prominent feature of these scenarios is that the chargino has a considerable lifetime and predominantly decays into a neutralino plus a low-momentum charged pion.
The low-momentum charged pion track is rarely reconstructed due to its  large displacement and a small number of interactions in the tracking system, therefore, a decaying chargino is typically recognized as a ``disappearing track'' that has few associated hits in the outer tracking volume.

The search for the direct production of meta-stable charginos is performed using events containing at least one large transverse momentum ($\pt$) jet (predominantly originating from initial-state radiation, and being used to trigger the signal event) with $\pt >90 \GeV$, $\MET>90 \GeV$ and $\Delta \phi_{\mathrm{min}}^{\mathrm{jet\mathchar`-}\met} > 1.5$, where $\Delta \phi_{\mathrm{min}}^{\mathrm{jet\mathchar`-}\met}$ indicates the smallest azimuthal separation between the missing transverse momentum and either of the two highest-$\pt$ jets.
Finding the disappearing-track signature suffers from charged hadrons interacting in the detector material, charged leptons losing much of their momenta due to scattering with material or large bremsstrahlung, and short-length tracks originating from a combination of wrong space-points and resulting in anomalously high values  $\pt$.
Their contributions to the signal search data sample are estimated using dedicated control data samples of each background component.

\begin{figure}[htb]
  \centering
  \subfigure[Track-$\pt$ spectrum]{
    \includegraphics[width=0.45\textwidth]{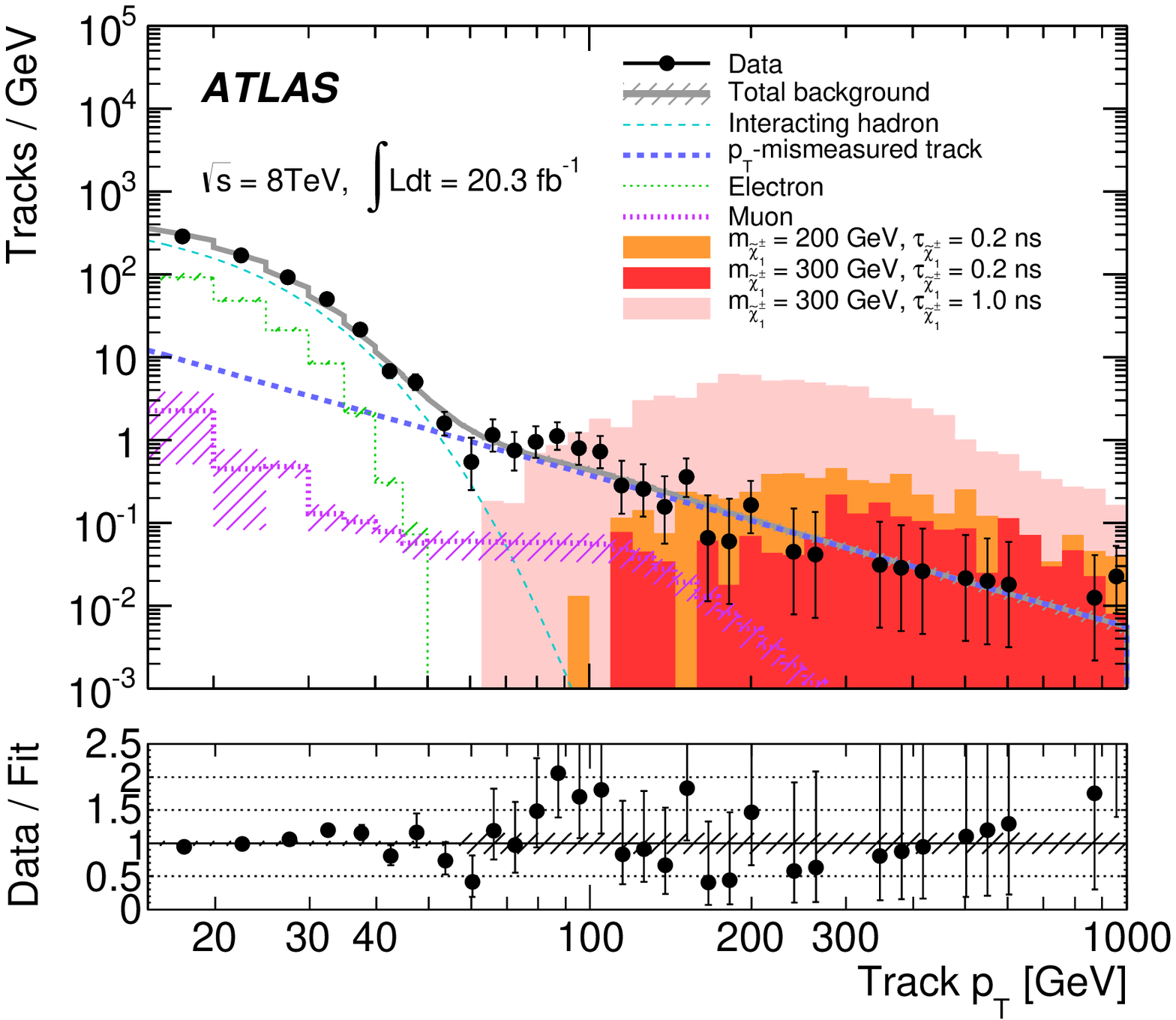}
  }
  \subfigure[The constraint on the allowed $\Delta m_{\tilde{\chi}_1}$--$m_{\chinoonepm}$ space]{
    \includegraphics[width=0.45\textwidth]{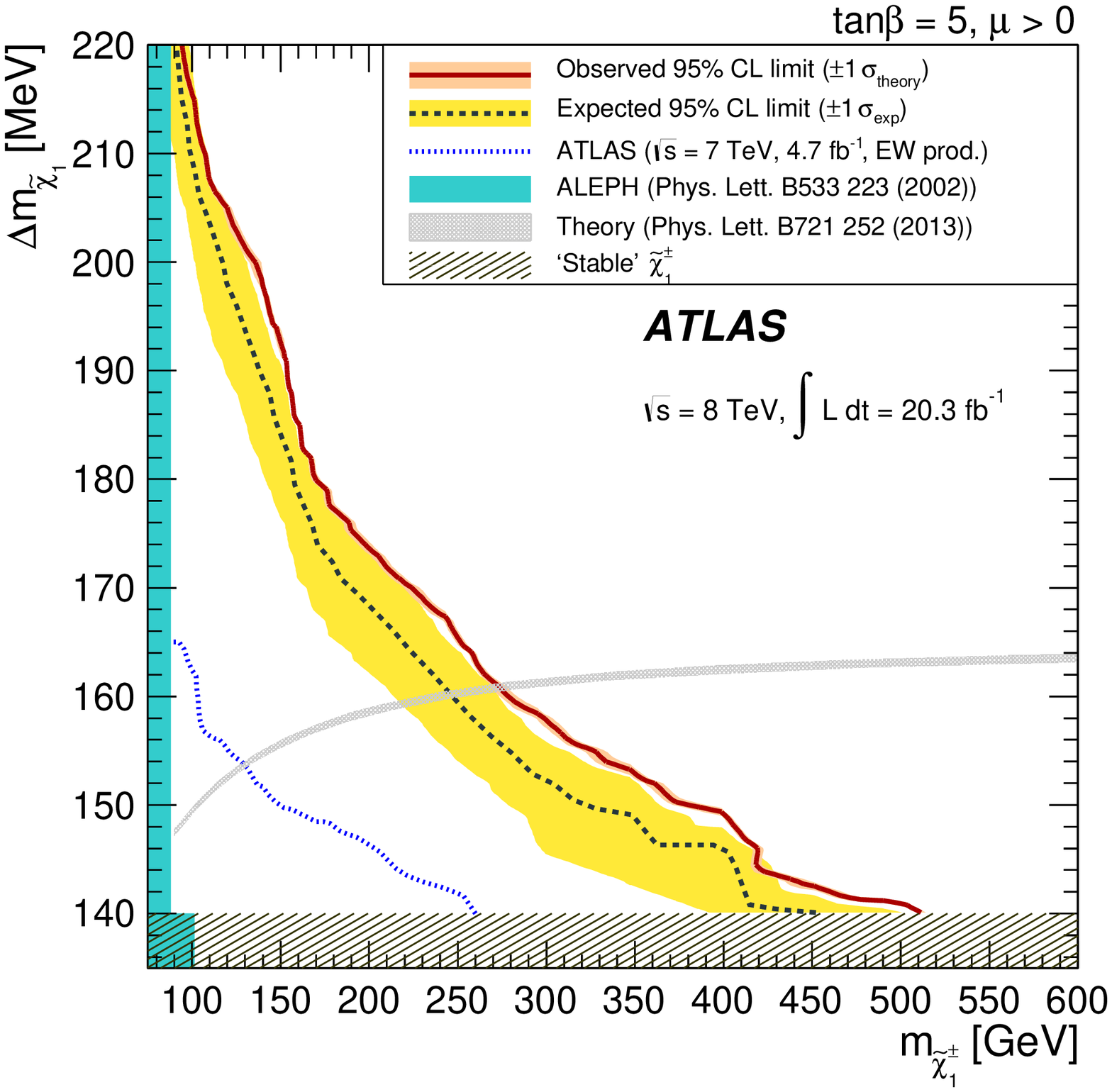}
  }
  \caption{The $\pt$ spectrum of disappearing-track candidates(a) and the constraint on the allowed $\Delta m_{\tilde{\chi}_1}$--$m_{\chinoonepm}$ space in the minimal AMSB scenarios(b).
  The search for charginos with long lifetimes, as indicated by the shaded region, is not covered by the analysis.
  }
  \label{fig:figure2}
\end{figure}

Figure~\ref{fig:figure2}(a) shows the  resultant $\pt$ spectrum of candidates tracks.
In the absence of a signal, constraints are set on the $\Delta m_{\tilde{\chi}_1}$--$m_{\chinoonepm}$ parameter space of the minimal AMSB model, as shown in Fig.~\ref{fig:figure2}(b), where $m_{\chinoonepm}$ is  the lightest chargino mass.
The theoretical prediction of $\Delta m_{\tilde{\chi}_1}$ for wino-like lightest chargino and neutralino states  is also indicated.
The limit that excludes charginos of $m_{\chinoonepm}< 270 \GeV$ (corresponding $\Delta m_{\tilde{\chi}_1}$ and the chargino lifetime being $\sim 160~\MeV$ and $\sim 0.2~\mathrm{ns}$, respectively) at 95\% Confidence Level (CL) is set.

 \subsection{Heavy stable charged particles}
Heavy stable charged particles, traversing the ATLAS detector, are identified by their velocities significantly lower than the speed of light and resulting large ionization losses.
The search utilizes the $\beta$ measurements based on the time-of-flight in the muon spectrometer and calorimeters:  the particle mass is calculated using the formulation $p/\beta\gamma$, where $p$ is the momentum of the charged particle track.
A high resolution on $\beta$ is achieved by a calibration using a data sample of muons from $Z$ decays.
The reconstructed mass is finally obtained by averaging the two $\beta$ measurements in different detector elements.
Background tracks dominantly originate from muons with mis-measured $\beta$; the mass spectrum of these tracks are estimated using a convolution of momentum and $\beta$ distributions of the signal search sample, relying on the fact that these two kinematic variables are fully uncorrelated for the background tracks.

\begin{figure}[htb]
  \centering
  \subfigure[The reconstructed mass distribution]{
    \includegraphics[width=0.45\textwidth, trim=0mm 20mm 0mm 0mm]{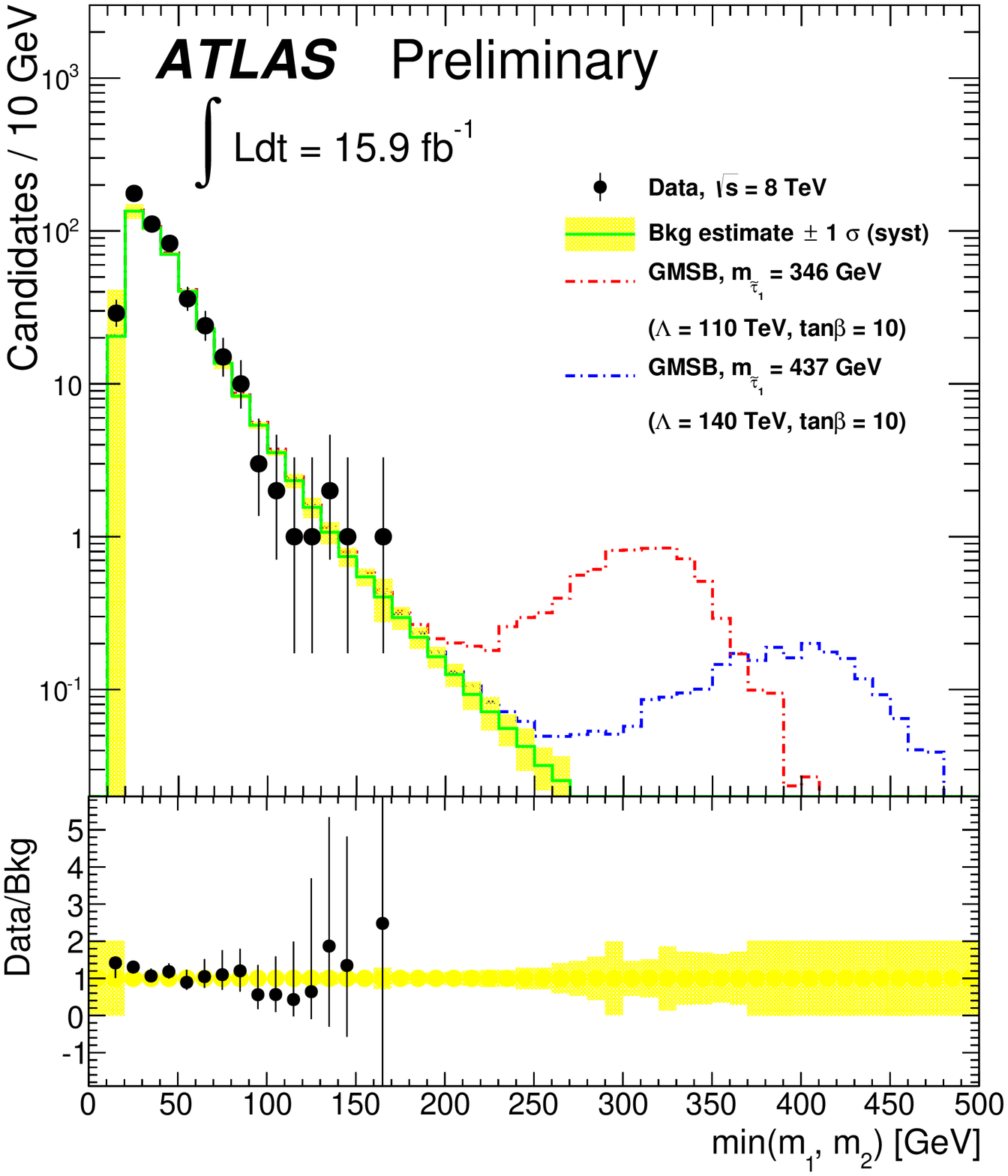}
  }
  \subfigure[Limits on the production cross section in GMSB scenarios]{
    \includegraphics[width=0.45\textwidth]{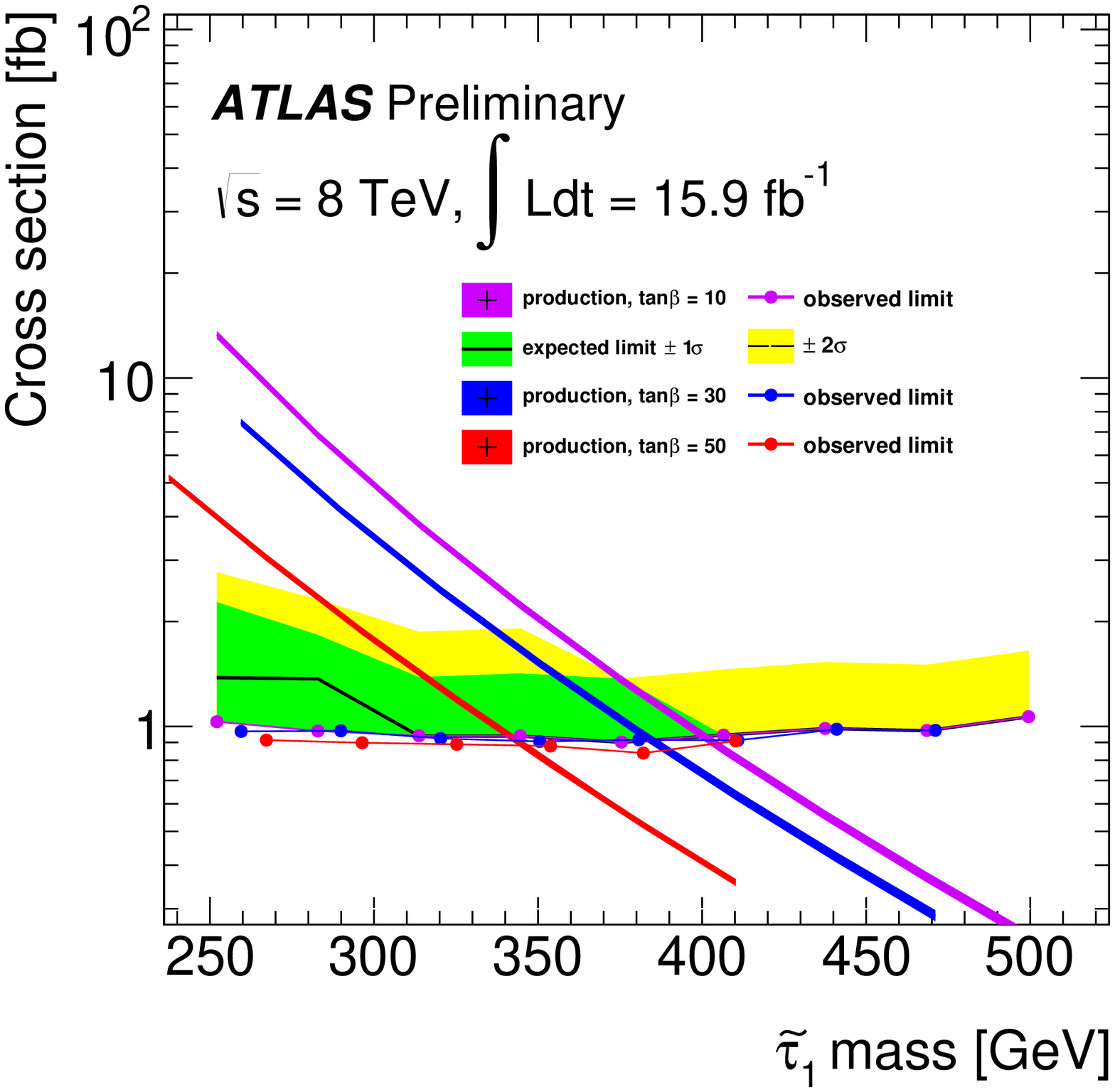}
  }
  \caption{ The reconstructed mass distribution(a) and limits on the production cross section as a function of stau mass in GMSB scenarios(b). }
  \label{fig:figure3}
\end{figure}

Figure~\ref{fig:figure3}(a) shows the resultant mass distribution of a signal search sample in which the candidate events are required to have two tracks with $\pt > 50 \GeV$ and $0.2<\beta<0.95$; no significant excess above the background expectation is found.
On the assumption that the stau is stable, the efficiency times acceptance for direct pair production is approximately 20\%, which gives  upper limits on the cross section of  $\sim 1~\mathrm{fb}$ at 95\% CL in the stau mass range from $250$ to $500~\GeV$.
The results are also interpreted in the context of gauge-mediated SUSY breaking (GMSB)~\cite{Giudice:1998bp} scenarios where a slepton is the next-to-lightest superpartner with a significantly long lifetime, as shown in Fig.~\ref{fig:figure3}(b).

\subsection{Displaced decays of heavy particles}
\begin{figure}[htb]
  \centering
  \subfigure[Signal diagram]{
    \includegraphics[width=0.30\textwidth]{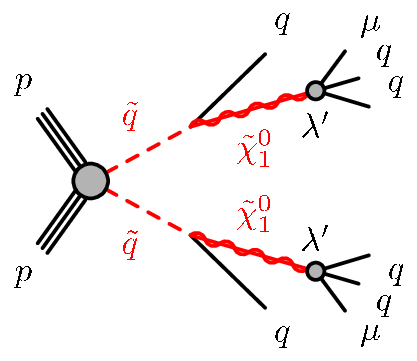}
  }\\
  \subfigure[$m_{\mathrm{DV}}$ vs. $N_{\mathrm{trk}}$]{
    \includegraphics[width=0.40\textwidth]{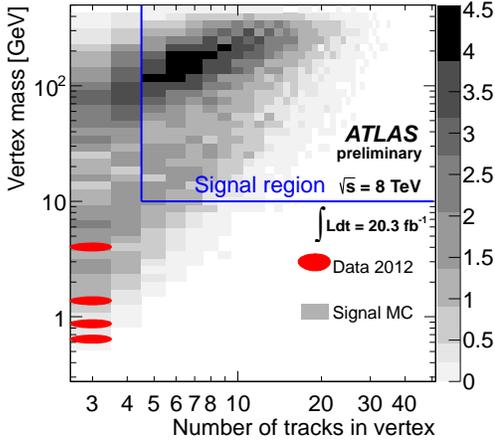}
  }
  \subfigure[Limits on the production cross section vs. neutralino lifetime]{
    \includegraphics[width=0.40\textwidth]{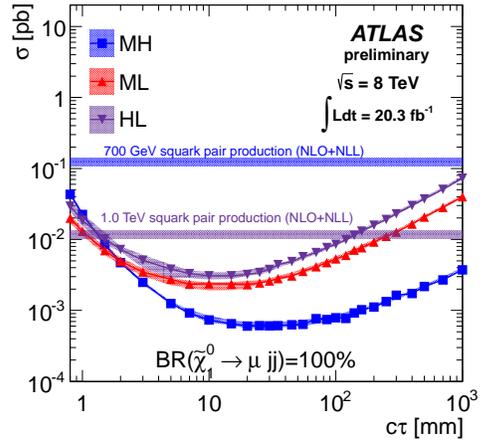}
  }
  \caption{Diagram of the signal model considered(a), the $m_{\mathrm{DV}}$--$N_{\mathrm{trk}}$ distribution(b), and the upper limit on the squark-pair production cross section as a function of  the neutralino lifetime for different combinations of squark and neutralino masses(c). The labels in the plot indicate the squark neutralino masses in GeV: 700--494 (MH), 700--108 (ML),  and 1000--108 (HL). The branching ratio of the decay chain from squark to neutralino to muon-plus-jets is set to 100\%.}
  \label{fig:figure4}
\end{figure}

One distinguishing signature of a long-lived particle is a significant displacement of its decay vertex from the $pp$ interaction point.
The ATLAS utilizes an enhanced track reconstruction to efficiently identify displaced vertices originating from heavy decaying particles with the lifetime range from $\mathcal{O}(10^{-12})$ to $\mathcal{O}(10^{-9})$ seconds. 
The present search is performed using events with a multi-track displaced vertex that contains a high-$\pt$ muon, and its result is interpreted in the context of a $R$-parity violating scenario in which non-zero value of $\lambda^{\prime}_{2ij}$ allows the lightest neutralino to decay into a muon and jets, as shown in Fig.~\ref{fig:figure4}(a).
Candidate events are selected with the requirements of a high-$\pt$ muon and a displaced vertex with $3 < r_{\mathrm{DV}} < 180~\mathrm{mm}$, where $r_{\mathrm{DV}}$ is the transverse distance of the vertex.
In order to suppress background events containing displaced verticies originating from secondary interactions, only vertices located outside the material of the inner detector are considered.
The vertex is finally required to have a mass ($m_{\mathrm{DV}}$) above $10~\GeV$ and a charged multiplicity ($N_{\mathrm{trk}}$) larger than 4.
Figure~\ref{fig:figure4}(b) shows the resultant $m_{\mathrm{DV}}$--$N_{\mathrm{trk}}$ two-dimensional distribution.
No event is found in the signal region while the expected number of background events (predominantly originating from random combinations of tracks and hadron interactions with air molecules) is $0.02 \pm 0.02$.

Based on this result,  an upper limit of 0.14 fb is set on the visible cross section at 95\% CL.
The corresponding 95\% CL upper limits on the production cross section is also set for some squark and neutralino masses as a function of the neutralino lifetime, as shown in Fig.~\ref{fig:figure4}(c).
A number of final states containing a displaced vertex are also being analyzed, which may give sensitivity to various representative models containing heavy long-lived particles such as split SUSY, GMSB, and various $R$-parity violating couplings.

\section{Conclusions}
Heavy long-lived particles are predicted by various SM extensions  and a number of searches  have been performed in the ATLAS experiment.
Although no evidence of signal is found to date, signatures of long-lived particles deserve to be continuously investigated with coming data of higher LHC beam energies as they could appear in SUSY parameter space that is not under the current experimental and phenomenological constraints.

\Acknowledgements
The author would thank all the members of the ATLAS collaboration and the conference organizers for the opportunity of presenting this work.

\end{document}